\documentclass[10pt,conference]{IEEEtran} 

\IEEEoverridecommandlockouts

\usepackage{url}
\usepackage{colortbl}

\usepackage{balance}
\usepackage{xspace}

\usepackage{graphicx}
\usepackage{verbatim}

\usepackage{bold-extra}
\usepackage{amsmath}
\usepackage{multirow}
\usepackage{listings}
\usepackage{boxedminipage}
\usepackage{soul}
\usepackage{enumitem}
\usepackage[normalem]{ulem}
\setlist{nolistsep,leftmargin=.5cm}
\usepackage{booktabs}
\usepackage{tabularx}
\usepackage{svg}
\usepackage{calc}
\usepackage{float}
\usepackage{multirow}
\usepackage[normalem]{ulem}
\useunder{\uline}{\ul}{}

\usepackage[most]{tcolorbox}
\definecolor{MidnightBlue}{HTML}{006895}
\definecolor{BoxesBlue}{HTML}{DEECFF}
\definecolor{BoxesYellow}{HTML}{FFF2CC}
\definecolor{StateGreen}{HTML}{91C788}
\definecolor{StateRed}{HTML}{FF8080}
\definecolor{ArrowGreen}{HTML}{61B15A}
\definecolor{ArrowViolet}{HTML}{BA94D1}
\usepackage{caption}
\usepackage{microtype} 

\usepackage{cite}

\usepackage{newtxtext,newtxmath}
\usepackage{algorithmic}
\usepackage{graphicx}
\usepackage{textcomp}
\usepackage{xcolor}

\usepackage[hidelinks]{hyperref}
\usepackage{ifthen}
\usepackage{wrapfig}
\usepackage{subcaption}
\usepackage{cleveref} 
\usepackage{enumitem}
\usepackage{indentfirst}
\usepackage[T1]{fontenc}

\newboolean{showcomments}
\setboolean{showcomments}{false}

\newboolean{showboxes}
\setboolean{showboxes}{true}

\usepackage{adjustbox}
\usepackage{totcount}

\ifthenelse{\boolean{showcomments}}

\makeatletter

\newcommand{\CrashScope}{{\sc CrashScope}\xspace}

\newcommand{\EulerC}{{\sc Euler}\xspace}

\newcommand{\Fusion}{{\sc Fusion}\xspace}

\newcommand{\ebug}{{\sc EBug}\xspace}
\newcommand{\burt}{{\sc Burt}\xspace}

\newcommand{\tool}{{\sc AstroBR}\xspace}

\newcommand{\toolTitle}{{\sc languAge underStanding and assessmenT of the steps to ReprOduce in Bug Reports}\xspace}

\definecolor{mycolor}{RGB}{204,204,204}

\newcommand*\circled[1]{\tikz[baseline=(char.base)]{\small{\textbf{
      \node[shape=circle,fill=mycolor,draw=black, inner sep=0.75pt] (char) {\textcolor{black}{#1}};}}}}

\newcounter{myboxcounter}

\setboolean{showcomments}{true}

\ifthenelse{\boolean{showcomments}}
  {\newcommand{\nb}[2]{
    \fbox{\bfseries\sffamily\scriptsize#1}
    {\sf\small$\blacktriangleright$\textit{#2}$\blacktriangleleft$}
   }
   
  }
  {\newcommand{\nb}[2]{}
   
  }

\newcommand{\ie}{\textit{i.e.,}\xspace}
\newcommand{\eg}{\textit{e.g.,}\xspace}
\newcommand{\etc}{\textit{etc.}\xspace}
\newcommand{\etal}{\textit{et al.}\xspace}

\newcommand\rev[1]{{\color{black}{#1}}}

\begin{document}

\title{Combining Language and App UI Analysis for the Automated Assessment of Bug Reproduction Steps
\thanks{The first two authors contributed equally to this work.}
}

\author{
	
	\IEEEauthorblockN{Junayed Mahmud\IEEEauthorrefmark{1}, Antu Saha\IEEEauthorrefmark{2}, Oscar Chaparro\IEEEauthorrefmark{2}, Kevin Moran\IEEEauthorrefmark{1}, Andrian Marcus\IEEEauthorrefmark{3}}
    \IEEEauthorblockA{\IEEEauthorrefmark{1}\textit{University of Central Florida (USA)}, 
		    \IEEEauthorrefmark{2}\textit{William \& Mary (USA)},
		    \IEEEauthorrefmark{3}\textit{George Mason University (USA)}
		    \\ \href{mailto:}{junayed.mahmud@ucf.edu}, \href{mailto:}{asaha02@wm.edu}, \href{mailto:}{oscarch@wm.edu},
		    \href{mailto:}{kpmoran@ucf.edu},
		    \href{mailto:}{amarcus7@gmu.edu}
    }
}

\maketitle

\thispagestyle{plain}
\pagestyle{plain}

\begin{abstract}
Bug reports are essential for developers to confirm software problems, investigate their causes, and validate fixes. Unfortunately, reports often miss important information or are written unclearly, which can cause delays, increased issue resolution effort, or even the inability to solve issues. One of the most common components of reports that are problematic is the steps to reproduce the bug(s) (S2Rs), which are essential to replicate the described program failures and reason about fixes. Given the proclivity for deficiencies in reported S2Rs, prior work has proposed techniques that assist reporters in writing or assessing the quality of S2Rs. However, automated understanding of S2Rs is challenging, and requires linking nuanced natural language phrases with specific, semantically related program information. Prior techniques often struggle to form such language $\leftrightarrow$ program connections -- due to issues in language variability and limitations of information gleaned from program analyses. 

To more effectively tackle the problem of S2R quality annotation, we propose a new technique called \tool, which leverages the language understanding capabilities of LLMs to identify and extract the S2Rs from bug reports and \rev{map} them to GUI interactions in a program state model derived via dynamic analysis. We compared \tool to a related state-of-the-art approach and we found that \tool annotates S2Rs 25.2\% better (in terms of F1 score) than the baseline. 
Additionally, \tool suggests more accurate missing S2Rs than the baseline (by 71.4\% in terms of F1 score). 
\looseness=-1
\end{abstract}


\section{Introduction}
\label{sec:intro}

End-users and developers frequently submit natural language bug descriptions through issue trackers in the form of bug reports.
These reports are essential in helping developers reproduce and understand the bugs, which in turn help in fixing them.
At the very least, a good bug report should describe the observed behavior (OB) of the app (\ie the buggy behavior), the expected behavior (EB) of the app (\ie the correct behavior), and the steps to reproduce the bug (S2Rs)~\cite{Zimmermann2010, Bettenburg2008GoodBR}.
Among these, the S2Rs are arguably the most important in reproducing the reported bug, an essential step in confirming the presence of the bug.

In GUI-based applications, reproducing a bug requires exercising a series of interactions via the Graphical User Interface~(GUI), as described by the S2Rs.
A developer (or a tool) trying to replicate a bug needs to understand and extract from each S2R description the user action (a click, swipe, \etc) and the GUI component the action is applied to (a button, menu, check box, \etc). 
This is often challenging, as end-users often use their own language and understanding of the app when describing the S2Rs, which may differ from that of the developers.
Incorrect or ambiguous S2R descriptions and missing S2Rs hinder developers' ability to understand the bug and lead to non-reproducible bugs \cite{ErfaniJoorabchi2014}, delays in bug fixes \cite{Guo2010, Zimmermann2012}, unresolved bugs \cite{Guo2010}, and even reopening bugs due to incorrect fixes \cite{Zimmermann2012}.
\looseness=-1

To address the problem of low-quality S2R descriptions in bug reports, previous research focused on generating missing S2Rs \cite{Bo2024}, providing quality feedback to bug reporters \cite{Chaparro2019}, automatically reproducing the bug reports  \cite{Fazzini2018}, or facilitating interactive bug reporting~\cite{song2022burt,Moran2015,song2020bee}.
A common issue shared by several of these approaches is related to difficulties in \rev{mapping} low-quality S2R sentences to elements of the GUI, stemming from the limitations of traditional natural language processing techniques.
\looseness=-1

In this paper, we present \tool (\toolTitle), a novel approach for improving bug reports at reporting time, by providing quality feedback on S2Rs to the reporter. To do this, \tool constructs an application execution model comprising the application interactions via dynamic analysis. Then, for each S2R, it identifies the corresponding application interactions via traversal of an app execution model, guided by GPT-4 \cite{chatgpt}. During the traversal, it identifies the best path comprising interactions for the first to last S2R of the bug report. Leveraging the interaction path and the mapped interaction information for each S2R, \tool can assess the quality of the reported S2R as well as generate the potential steps that are not reported in the bug, but required to reproduce the bug (\ie\ missing steps). 
\looseness=-1

\begin{figure*}[t]
		\centering
		\includegraphics[width=\linewidth]{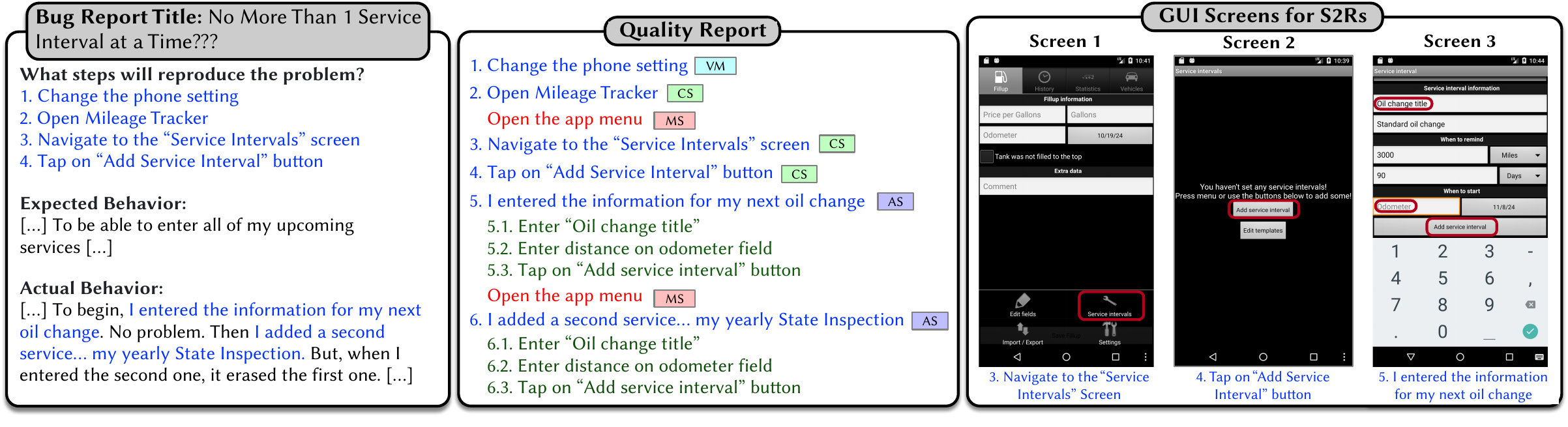}
		\caption{Bug Report Quality Annotations}
		\label{fig:bug-report}
		\vspace{-0.3cm}
\end{figure*}

Unlike previous work, \tool uses an LLM (GPT-4) for three different tasks, in three distinct ways. 
\textit{First}, it automatically extracts S2R sentences in a natural language bug report, framing the task as a text classification problem.
For this task, we evaluated three prompt templates, based on three prompting strategies (\ie zero-shot, few-shot, and chain-of-thought),  using a development set of 54 bug reports.
\textit{Second}, it extracts individual user actions and the GUI components interacted with from the S2R sentences, framing the task as a phrase extraction problem.
For this task, we evaluated three additional prompt templates, based on the three prompting strategies.
\textit{Third}, it maps the extracted actions and GUI components to elements of an app execution model, framing the task as a guided graph exploration problem.
For this task, we evaluated six prompt templates, based on the three prompting strategies.
GPT-4 is used to guide the systematic and efficient exploration of the execution model.
During this mapping process, \tool identifies problems with the S2R sentences (\eg ambiguous descriptions, vocabulary mismatches, or missing steps) and generates a quality report with annotations reflecting these issues.
When S2Rs are missing, \tool also generates the missing steps and includes them in a quality report.
\looseness=-1

We compared the performance of \tool with a recent state-of-the-art technique, \EulerC \cite{Chaparro2019}, utilizing a test dataset consisting of 21 bug reports having 73 S2R sentences, from five Android applications. 
\tool achieves better results in generating quality annotations (by 25.2\% F1 score) and identifying missing S2Rs (by 71.4\% F1 score).

In summary, this paper makes the following contributions:
\begin{itemize}
	\item An approach that integrates graph-based dynamic analysis and LLMs (\ie GPT-4) to automatically identify and extract S2Rs from the bug report, assess their quality, and generate missing steps. 
	\item A ground truth dataset containing the identified and extracted S2Rs, quality annotations, and missing steps to evaluate \tool with the existing baselines. This dataset contains 75 annotated bug reports.
	\item A replication package~\cite{package,doi} containing all the dataset and source code to replicate and validate our results.
\end{itemize}
\looseness=-1


\section{Quality Model for Reproduction Steps}
\label{sec:quality_model}

In this paper, we adopt the quality model proposed by Chaparro \etal \cite{Chaparro2019}, with the following quality categories for the steps to reproduce the bug (S2Rs) in a bug report: 
\begin{itemize}
	\item Correct step (\textbf{CS}): the step corresponds to a specific interaction and GUI component on the application.
	\item Ambiguous Step (\textbf{AS}): the step corresponds to multiple interactions on GUI components on the application.
	\item Vocabulary Mismatch (\textbf{VM}): the step does not correspond to any interactions or GUI components on the application due to misaligned terminology.
	\item Missing Steps~(\textbf{MS}): interactions that are required to replicate the bug, but not reported in the bug report.
\end{itemize}

We illustrate the definitions with an example in Figure \ref{fig:bug-report}. 
The bug report presented in the figure comprises six S2Rs, each annotated with the above categories. \noindent\circled{1} The first S2R is \textit{"Change \rev{the} phone setting"}, which does not represent any interactions in the app. 
Therefore, this S2R is annotated as \textbf{VM}.
\noindent\circled{2} The second S2R contains only one individual S2R, \textit{"Open \rev{Mileage Tracker}"}, representing only one app interaction. Therefore, this S2R is annotated as \textbf{CS}. \noindent\circled{3} The third S2R, \textit{"Navigate to the `Service Intervals' screen"}, does not immediately follow after the second step. There is a required intermediate step, \textit{"Open the app menu"}, which must be performed by tapping the "three dots" button in the bottom left menu bar of Screen 1.
Therefore, this missing step is included in the quality report and annotated as \textbf{MS}. 
With this missing step added, the third reported S2R requires a single interaction that can be reliably mapped to the GUI, \ie\ performing a \textit{click} operation on the "Service Interval" button on Screen~1. Therefore, it is categorized as \textbf{CS}. 
\noindent\circled{4} \textit{"Tap on `Add Service Interval'} requires only one interaction in the GUI, \ie\ performing a \textit{click} operation on "Add Service Interval" component on Screen~2, and hence, is annotated as \textbf{CS}. 
\noindent\circled{5} The fifth S2R, \textit{"I entered the information for my next
oil change"}, requires multiple operations. At first, a user has to enter the \textit{"Oil change title"} by performing a \textit{type} operation on the "title" text field at the top of Screen 3. The individual S2R for this interaction is \textit{"Enter Oil change title"}. Secondly, s/he has to enter a value in the "Odometer" text field on Screen 3 by performing a \textit{type} operation which implies an individual S2R: \textit{"Enter distance on odometer field"}. Finally, s/he has to perform a \textit{click} operation on the "Add Service Interval" button on Screen 3. This interaction represents the individual S2R: \textit{"Tap on Add service interval button"}. As three interactions are required to complete the fifth step in the bug report, it is labeled as \textbf{AS}.
\noindent\circled{6} To execute the sixth S2R, \textit{"I added a second service my yearly State Inspection"}, there is another step missing, \textit{"Open the app menu"}, and it is labeled as \textbf{MS}. 
Moreover, the sixth step requires the same three individual S2Rs as the fifth step and annotated as \textbf{AS}. In the next section, we explain how this quality model can be automatically applied to bug reports.


\section{\tool: Automated S2R Quality Assessment}
\label{sec:approach}
\begin{figure*}[t]
		\vspace{-2em}
		\centering
		\includegraphics[width=1\linewidth]{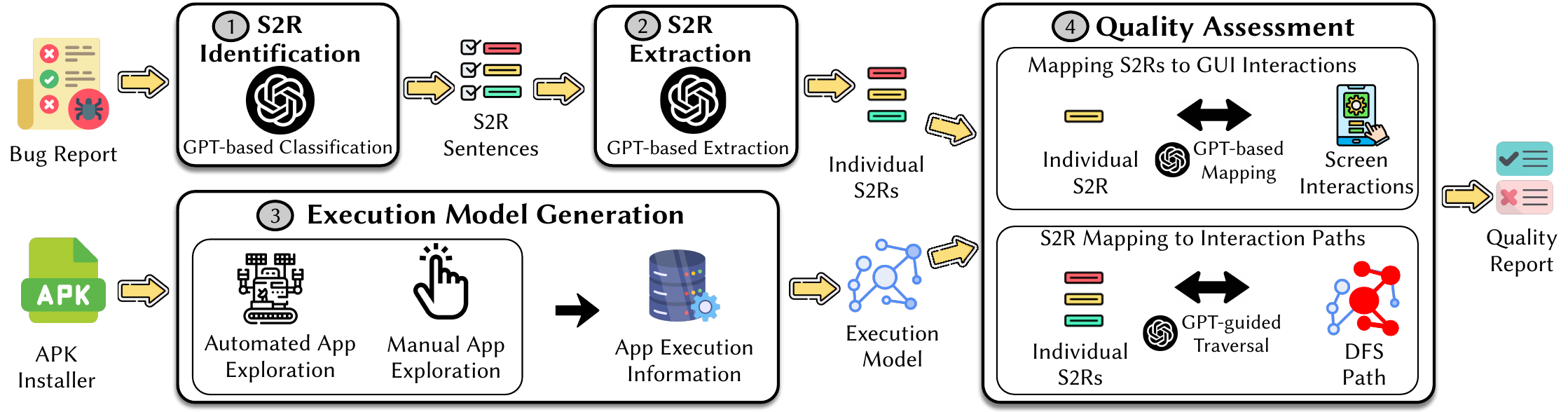}
		\caption{{The \tool Approach}}
		\label{fig:approach}
\end{figure*}

This section presents \tool, an automated approach that leverages an LLM and a graph-based app execution model to assess the quality of the steps to reproduce (S2Rs) in textual bug reports.  
\tool identifies, extracts, and processes the S2Rs from a bug report to detect which ones are correct, ambiguous, missing, or phrased using language that does not correspond to a target app, according to the quality model described in \Cref{sec:quality_model}.
 \tool generates a quality report with annotations that provide feedback to the reporter about problematic S2Rs and includes generated missing S2Rs.
\tool has four main components, as illustrated in \Cref{fig:approach}:
\begin{enumerate}
	\item \textbf{S2R sentence identification}: \tool identifies the sentences that describe any S2Rs (\Cref{sec:identification-phase}).
	\item \textbf{Individual S2R extraction}: \tool extracts phrases describing individual S2Rs from S2R sentences (\Cref{sec:indiv-s2rs-approach}).
	\item \textbf{App execution model generation}: \tool builds a graph-based model using automated and manual app execution (\Cref{sec:execution-model}).  
	\item \textbf{S2R quality assessment}: \tool maps individual S2Rs to GUI-level interactions captured in the app execution model, providing feedback about high- and low-quality S2Rs as well as missing steps in a quality report (\Cref{sec:quality-assessment-annotations}).
\end{enumerate}

We leverage the language processing capabilities of LLMs \rev{(\ie\ GPT-4)} across the three phases, integrating these with GUI-level dynamic app analysis to assess S2R quality. \rev{The selection of GPT-4 as the LLM was based on its demonstrated effectiveness in language and bug understanding tasks, including bug reproduction~\cite{Feng2024} and analysis~\cite{Bo2024}.}
In the remainder of this section, we detail \tool's components or phases.

\subsection{S2R Sentence Identification Phase}
\label{sec:identification-phase}

\tool automatically identifies sentences that describe any steps to reproduce (S2R) in the bug report (see the blue sentences in Fig. \ref{fig:bug-report}). 
This is necessary as the bug report typically includes other content, notably the observed (OB) and expected app behaviors (EB). 
We formulate this task as a text classification task, using LLMs. 
\tool decomposes the bug report into a list of sentences and asks the LLM to identify which of these sentences describe any S2Rs. 
Sentence parsing is done using the Stanford CoreNLP toolkit~\cite{manning2014stanford} and heuristics. 
We experimented with three types of prompts, each one providing a different context to facilitate the task for the LLM (\eg the definition of S2Rs and guidelines on how to distinguish them from other content like the OB and EB). 
\Cref{sec:prompt_development} describes the process we followed to develop and evaluate these prompts.
\looseness=-1

\subsection{Individual S2R Extraction Phase}
\label{sec:indiv-s2rs-approach}

After identifying S2R sentences, \tool asks the LLM to extract the individual S2Rs from these sentences in a particular format (described below). 
Individual S2Rs are phrases that describe a single, atomic interaction with the app. 
Individual S2R extraction is needed because S2R sentences may describe multiple interactions with the app together with content such as the OB (\eg ``I opened the app and clicked on the Start button'' or ``The app crashes if the user checks the Angle Box''). 
In addition, different S2R sentences may describe the same interaction (\eg "... the user checks the Angle Box" and "give the Exercise a name and check the Angle Box"). 
\tool resolves this redundancy by asking the LLM to provide only one S2R among all extracted individual S2Rs that describe the same interaction.
\looseness=-1

The output of this phase is a list of individual S2Rs extracted in the order they appear in the sentences from left to right and top to bottom. 
\tool asks the LLM to represent the individual S2Rs in the following format: 
\texttt{\small[action][object][preposition][object2]}. 
The \texttt{\small[action]} is a verb associated with the app interaction (tap, long tap, enter, \etc). 
The \texttt{\small[object]} is the GUI component upon which the action is performed. 
The \texttt{\small[object2]} is additional information related to the object connected by a \texttt{\small[preposition]}. 
For example, the S2R ``Click any button on this page" is formatted as \texttt{\small[Click] [any button] [on] [this page]}.

We designed and evaluated three prompt types to extract individual S2Rs via GPT-4. 
Each prompt implements a different approach, providing different contexts about the task (\eg examples that illustrate how to accomplish the task). 
\Cref{sec:prompt_development} details the prompts and the process we followed to design and evaluate them.
\looseness=-1
 
\subsection{App Execution Model Generation Phase}
\label{sec:execution-model}

\tool's quality assessment relies on mapping individual S2Rs to interactions that can be executed on the app to replicate the reported bug. 
This requires collecting and representing possible user GUI interactions, for which we adapt graph-based representations and dynamic app execution strategies from prior work~\cite{song2022toward,saha2024toward}.

\tool creates an app execution model represented as a directed graph, $G = (V, E)$, where $V$ represents the set of unique GUI screens for an app, and $E$ represents the set of unique interactions that users can perform on the GUI components of the screens. 
A GUI screen (\ie node) is represented as a hierarchy of the GUI components and layouts. 
Two GUI screens with different GUI component hierarchies are considered distinct graph nodes. 
Each interaction (\ie edge) in $E$ is represented by a unique tuple in the form of ($v_x, v_y, e, c)$, where $c$ is a GUI component of screen $v_x$ and  $e$ in an action (tap, type, \etc) performed on $c$, resulting in a transition to another screen $v_y$. 
Each edge contains additional interaction metadata such as the interacted GUI component type, ID, text (\ie label), and description. 

To build the execution model for an app, \tool parses GUI interaction traces collected from automated app exploration and manual app usage.  \tool executes an adapted version of the \CrashScope tool~\cite{Moran2016, Moran2017}, which implements multiple
automated exploration strategies to interact with the UI components of app screens, trying to exercise as many app screens and GUI components as possible.
In the process, \CrashScope collects app screenshots and XML-based GUI hierarchies and metadata for the exercised app UI screens and components. 
As \CrashScope may \rev{fail} to interact with certain GUI screens and components that app users would normally interact with, \tool can also make use of interaction data collected from manual app usage and testing. 
In this paper, for the development set, we used the set of traces collected by Saha \etal~\cite{saha2024toward} which consists of 10-12 manually recorded feature interaction traces for each of the 5 test applications. For the prompt development dataset, two authors collected the same number of traces for each of the 31 apps. These recordings include all the app GUI interactions starting from launching the application to the last step related to carrying out an application feature (more details of this process, used for prompt development and evaluation, are found in \Cref{sec:dev_dataset}). 
In practical applications of \tool, manual executions can be collected in several ways.
For example, developers can enable user monitoring features in the app and perform record-and-replay during in-house or crowd-sourced app testing~\cite{du2022semcluster}. 
\tool parses the interaction traces generated by \CrashScope and the traces collected during app usage/testing to build the graph, according to the graph format we previously described in this section (details found in \Cref{sec:dev_dataset}). 
 
\subsection{S2R Quality Assessment Phase}
\label{sec:quality-assessment-annotations}

The app execution model captures possible interaction sequences that a user could perform when using or testing an app as paths in the graph. 
To assess the quality of the S2Rs, \tool attempts to map each individual S2R to interactions (\ie edges) along these paths. 
To do so, \tool implements an LLM-guided depth-first-search (DFS) graph traversal 
to establish the correspondence between an individual S2R and interactions on a given screen. 

Any S2Rs that cannot be mapped to a graph interaction are labeled as having a Vocabulary Mismatch (\textbf{VM}).  
S2Rs that map to multiple interactions performed on a single screen (\ie a node) are labeled as Ambiguous Steps (\textbf{AS}). 
Those that map to single interactions within a sequence are labeled as Correct Steps (\textbf{CS}). 
Finally, for the mapped S2Rs that correspond to non-consecutive interactions spanning different screens in a path, additional interactions are required to connect them to form a complete path. 
These additional interactions are used to generate individual S2Rs that are labeled as Missing Steps~(\textbf{MS}) and used to fill in the "gaps" between the existing S2Rs.

\subsubsection{Mapping Individual S2Rs to Interactions on a Screen}
\label{sec:qualtiy_phase:mapping_single_screen}
Mapping an individual S2R (S2R, hereon)
to interactions on a given screen is supported by GPT-4. 
For a graph node (\ie a screen), \tool asks GPT-4 to identify which of the outgoing edges (\ie interactions) from that node correspond to the S2R. 
Both the S2R and graph interactions are represented textually: the S2R is extracted from the bug report, while each interaction is represented as a tuple of textual information (\eg the event description and the label of the interacted GUI component).  
We designed and evaluated a set of prompts using different prompting strategies to accomplish this mapping in a 2-step manner: a first prompt asks GPT-4 to return a yes/no answer on whether an individual S2R maps to the interactions of a given screen and if the answer is yes, a second prompt asks GPT-4 to return the list of corresponding interactions. The methodology used to develop and evaluate the prompts is detailed in \Cref{sec:prompt_development_methodology}. 

\subsubsection{Graph Traversal and S2R Mapping to Interaction Paths}
\label{sec:graph-traversal}

To map all the S2Rs from a bug report to app interaction sequences, \tool implements an algorithm that traverses the graph in a depth-first-search (DFS) manner, aiming to \rev{map} the S2Rs to interactions along the DFS paths. 
When S2Rs map to non-consecutive interactions within a path, \tool connects these interactions by selecting the shortest path between the nodes where these interactions occur. 
Since multiple paths may map to the S2Rs, \tool selects the path with the most mapped S2Rs or the shortest path, if multiple paths have the same number of mapped S2Rs.

The DFS traversal of the graph is guided by the LLM-based mapping approach from \Cref{sec:qualtiy_phase:mapping_single_screen}, as only edges that map to S2Rs are traversed, avoiding the need to explore the entire graph. 
While none of the S2Rs can map to any interaction in the graph (in which case \tool would traverse the graph entirely), this scenario is expected to be rare, as we assume reporters would describe at least one S2R using the app’s vocabulary and the graph is as complete as possible, covering a broad range of screens and interactions.

\textbf{\textit{Algorithm Details.}} 
\tool's DFS-based graph traversal algorithm is recursive. 
It receives an S2R $s$ and graph node~$n$ as input, where $s$ is the first item in the S2R list $L$ (the bug report S2Rs). 
The algorithm returns either: the best DFS path $p$ (starting from $n$) that maps to a subset of S2Rs in $L$ (possibly including $s$), or no path if no S2Rs can be mapped. 

The traversal begins with the first S2R from the bug report and the starting node of the graph, which contains "\textit{open app}" interactions that navigate to the screens users usually see upon launching the application. 

The algorithm has two main logic branches:
\begin{enumerate}
	\item If S2R $s$ does not map to any of the outgoing interactions~$I$ from $n$, the algorithm recurses, attempting to map  $s$ on each node connected to $n$ by $I$.  If this traversal results in no DFS paths mapped to $s$ or following S2Rs in $L$, $s$ is labeled as having a Vocabulary Mismatch (\textbf{VM}), and the algorithm recurses with the next S2R in $L$ at the current node $n$. 
	This means that the S2R $s$ cannot be mapped to any node in the (sub)graph starting from $n$, then the algorithm attempts to map the next S2R.
	\item Conversely, if $s$ maps to interactions in $I$, the algorithm checks whether there are one or more mapped interactions. If there is a single interaction, $s$ is labeled as a Correct Step~(\textbf{CS}); if there are multiple, it is labeled as an Ambiguous Step (\textbf{AS}). 
	The algorithm then recurses with the next S2R in $L$ on each node connected to $n$ by only the mapped interactions from $I$. 
	Essentially, if the algorithm succeeds at mapping $s$ to interactions from $n$, then it proceeds with attempting to map the next S2R to the resulting nodes after navigating to the mapped interactions.
\end{enumerate}

It is possible that $s$ maps to interactions in $I$ (second branch above), but there are "gaps" between the previous mapped S2R and $s$: if their mapped interactions are not consecutive in the DFS path. 
If this is the case, the algorithm connects them by determining the shortest path between the involved nodes. 
The interactions used to connect the nodes are then labeled as Missing Steps (\textbf{MS}). 
Note that this shortest path may include interactions outside the DFS path, as we are not limiting the shortest path search to the DFS path alone. A shorter path may exist that bypasses parts of the DFS path.

After traversing a node with a given S2R (in either branch above), it is possible that when calling the algorithm recursively on a set of interactions (\ie when navigating down DFS paths), it returns multiple DFS paths mapped to the S2Rs. 
If this is the case, the algorithm selects the DFS path to return based on the following criteria: prioritizing the path with the most mapped S2Rs in $L$, or, if paths have the same number, choosing the shortest path.
\looseness=-1

The traversal continues until all S2Rs in $L$ have been exhausted or until none of the S2Rs are mapped to any DFS paths. 
If all S2Rs have been mapped, but there are still nodes along a DFS path, the algorithm does not proceed to check additional nodes down the current DFS path.
To prevent re-processing nodes and their interactions, the algorithm marks each (node, S2R) pair as visited before it processes the node and S2R.
\looseness=-1

\subsubsection{Quality Report Generation}

The returned DFS path contains interactions mapped to all or a subset of the S2Rs from the bug report. Each S2R is labeled as either a Correct Step (CS), Ambiguous Step (AS), or Vocabulary Mismatch Step (VM).  
In addition, interactions identified to fill in the "gaps" between S2Rs are labeled as Missing Steps (MS). 
For evaluation purposes, we also mark the corresponding S2Rs with missing steps as MS, so that we can perform a fine-grained analysis of results (more details found in Section \ref{sec:empirical_evaluation}).


\section{\tool's Prompt Development and Evaluation}
\label{sec:prompt_development}

This section describes how we developed and evaluated the LLM prompts for three distinct tasks: (i) S2R sentence identification, (ii) individual S2R extraction, and (iii) individual S2R mapping to app interactions. 
We adopted a rigorous, comprehensive, and data-driven approach in which we designed an initial prompt that was iteratively evaluated and refined into new prompts. 
Prompt development and evaluation followed a quantitative and qualitative methodology based on a set of Android app bug reports. 
Overall, we designed and evaluated 12 prompt templates across all three tasks.
\rev{To generate GPT-4 responses with the prompts for all tasks, we used a temperature of 0 to minimize randomness/non-determinism in the responses.}

\subsection{Development Dataset Construction}
\label{sec:dev_dataset}

We constructed a dataset of 54 bug reports and corresponding ground truth data, with manually identified S2R sentences, individual S2Rs, and interactions mapped to each S2R.

\subsubsection{Bug Report Collection}

We selected the 54 bug reports from the dataset released by Saha \etal~\cite{saha2024toward}, which contains reproducible mobile app bug reports from the AndroR2 dataset~\cite{wendland2021,Johnson2022}. 
These reports describe bugs for 31 Android apps of various domains (\eg web browsing, WiFi network diagnosis, and finance tracking). 
The reported bugs span different bug types, namely crashes (15 reports), output problems~(19), UI cosmetic issues (13), and navigation problems (7). 

\subsubsection{S2R Sentence Labeling}
\label{sec:identification_data_dev}

Two authors annotated the 1,031 sentences present in the bug reports as either S2R or non-S2R, following the S2R criteria and methodology defined by Chaparro \etal~\cite{Chaparro2017}. 
One author annotated each sentence, while the second author validated the annotations, recording disagreements and their rationale. 
The authors agreed on the annotations for 1,002 sentences (97.2\%, 0.91 Cohen's kappa~\cite{Cohen}), which represents near-perfect agreement. 
Disagreements were resolved via discussion. 
The most common reasons for disagreements were content misinterpretations and mistakes (\eg\ a sentence describing the observed behavior, not S2Rs). 
In total, the 54 bug reports contain 189 S2R sentences (3.5 per report on average), while the remaining 842 sentences describe non-S2R content.
\looseness=-1

\subsubsection{Individual S2Rs Extraction}
\label{sec:extraction_data_dev}
Two authors manually inspected the 189 S2R sentences to extract individual S2Rs (phrases describing a single interaction with the app). 
One author read and extracted the individual S2Rs in the format defined in \Cref{sec:indiv-s2rs-approach}. 
The extracted S2Rs were validated by a second author. 
They discussed disagreements to reach a consensus where needed. 
From the 189 S2R sentences, we extracted 246 individual S2Rs with an agreement rate of 97.6\%. 

\subsubsection{S2Rs to GUI Interaction Mapping}
\label{sec:qa_data_dev}

To create ground truth mappings between individual S2Rs and GUI app interactions, we first built the execution models (\ie graphs) for the 31 apps corresponding to the bug reports. 
To do so, we executed the \CrashScope tool~\cite{Moran2016} using the corresponding APKs (from the original dataset~\cite{saha2024toward,Johnson2022}) and a Pixel 2 Android emulator. 
We also used the manual interaction traces collected as part of Saha \etal's dataset~\cite{saha2024toward}. Both the \CrashScope and manual interaction traces consist of GUI-event execution traces and (video) screen captures showing the executed interactions. 
We used Song \etal's toolkit~\cite{song2022burt} to parse the traces and build the execution graphs.

Two authors manually inspected the execution data, graphs, and reproduction screen captures to map each S2R to graph nodes and interactions. 
One author first inspected this data to identify the GUI screen and target GUI component for each S2R. 
Then, the author identified the graph node corresponding to such screen, and within it, the interaction corresponding to the S2R. 
In the process, missing steps and the path that represented a minimal bug reproduction scenario were identified. 
A second author followed the same procedure to verify the interactions/nodes mapped to the S2Rs and the reproduction paths identified by the first author. 
Both authors discussed any disagreements, involving a third author where necessary.  
\looseness=-1

We applied the above methodology on a sample of 10 bug reports, in such a way that they spanned different bugs types, 
apps of different domains (9 apps), and S2R types (taps, types, \etc). 
The two authors created the ground truth for 46 individual S2Rs among 49 individual S2Rs for the 10 bug reports, agreeing on 43 S2Rs (agreement rate of 93.5\%). The excluded three individual S2Rs did not have corresponding app interactions in the execution model because they are performed outside the app (\eg\ \textit{"install the app"}), and hence, are not included in the graph.
Common reasons for disagreements were unclear individual S2Rs and misinterpretation of graph nodes/interactions. 
During the data creation process, we realized that it would take the two authors a prohibitive amount of effort to create the data for the remaining 44 bug reports. 
Therefore, we decided to focus on the S2R mapping prompt development using only the 10 bug reports and redirect our effort to curating the test data used for \tool's evaluation (see \Cref{sec:empirical_evaluation}). 

\subsection{Prompt Development Methodology}
\label{sec:prompt_development_methodology}

\begin{figure}
	\centering
	\includegraphics[width=\linewidth]{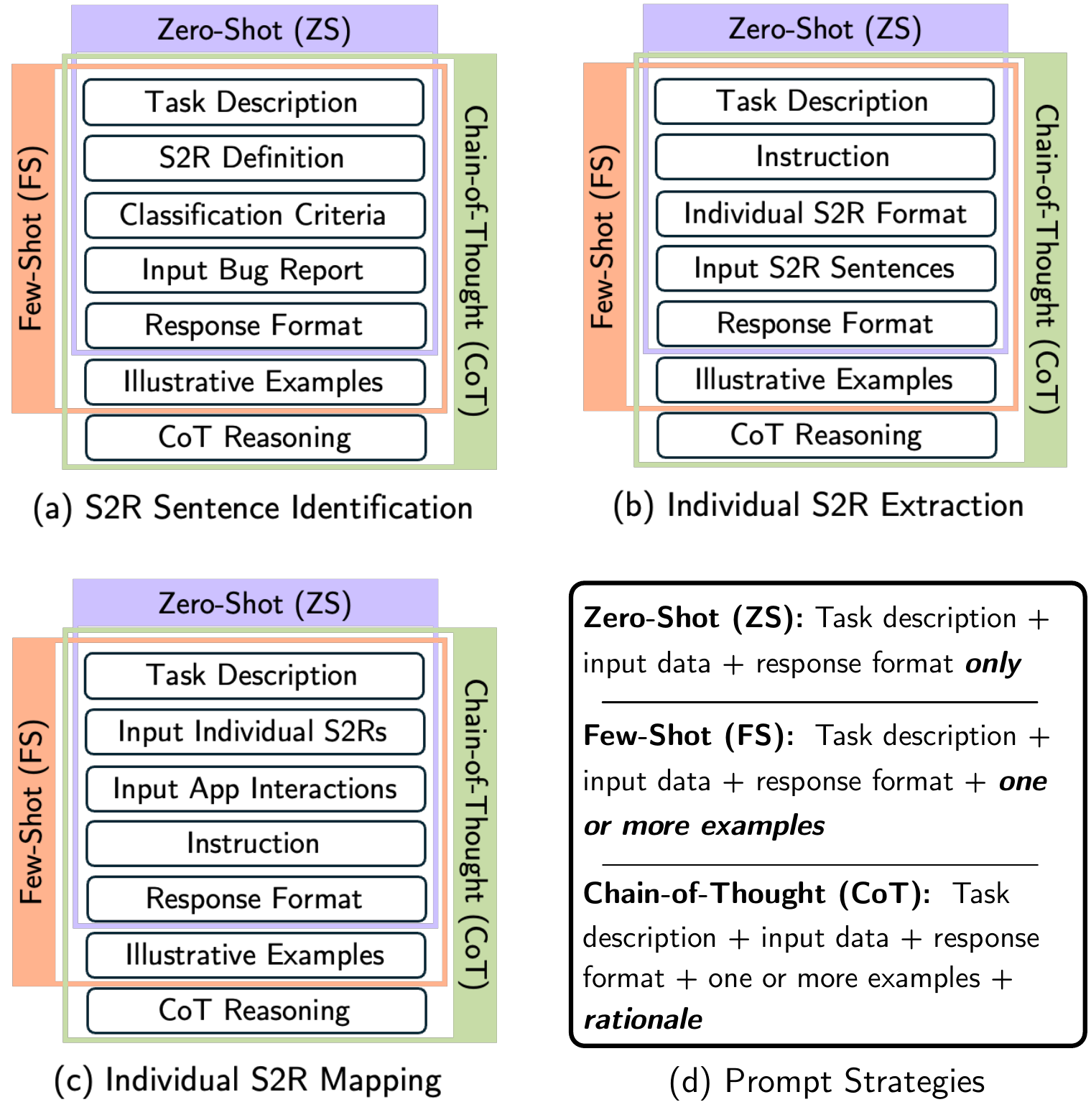}
	\caption{Structure of the Developed Prompts}
	\label{fig:prompt-structure}
\end{figure}

For each of three tasks where \tool uses GPT-4, our overall data-driven methodology used three prompting strategies, commonly used in software engineering research~\cite{hou2023large}: 
\begin{itemize}
	\item Zero Shot (ZS) prompting: starting from a base prompt template that includes the task description, input, and response format, we iteratively executed, evaluated, and refined the template until the performance plateaued. This involved computing performance metrics (precision, recall, and F1 score) against the ground truth, qualitatively analyzing false positives (FP) and negatives (FN), and adjusting the prompt to address those cases. For example, as S2R sentence identification is a classification task, two authors investigated the FP and FN of the GPT-4 responses to derive the classification criteria (\Cref{fig:prompt-structure}a) to better guide GPT-4 in the S2R sentence classification task.
    \rev{This process resulted in four versions of each type of prompt template. To determine if performance plateaued, we monitored the F1 score. For example, from version 3 to version 4 the F1 score decreased by 0.001 for the S2R identification task. Based on this minimal change, we selected version 3 as the optimal prompt for this phase.}
	\item Few Shot (FS) prompting: starting from the obtained ZS template, we created a base FS template containing positive and negative examples selected from the remaining bugs of Saha \etal's dataset~\cite{saha2024toward} and the expected output. The example bug reports are representative of each task and selected based on certain criteria, \eg\ various bug types (crash, output, \etc), and bug reports with different wordings and structures. We iteratively executed, evaluated, and refined the template until the performance no longer improved, in the same way we did it in ZS prompting.
	\item Chain of Thought (CoT) prompting: starting from the obtained FS template, we created a base CoT template containing explanations for \rev{the outcome of the positive and negative examples. The explanation for the outcome was designed by two authors after discussion and reaching a consensus.} We iteratively executed, evaluated, and refined the template until the performance plateaued, in the same way we did it in ZS and FS prompting.
\end{itemize}

This methodology resulted in three prompt templates (one from each prompting strategy) for S2R identification and three templates for individual S2R extraction. 
For S2R mapping, since we defined mapping as a 2-step task, we designed two prompts for each strategy, resulting in six prompts. 
The 2-step task consisted of first asking GPT-4 to return a yes/no answer on whether an individual S2R maps to the interactions of a given screen and if the answer is yes, asking GPT-4 to return the list of interactions that the S2R maps to. 
Our tests revealed that this approach led to less noisy answers from GPT-4, compared to executing only the second step. 
In total, we developed 12 prompt templates. To help visualize our prompt templates, \Cref{fig:prompt-structure} illustrates the various components associated with the prompts for each task---our detailed templates are found in our replication package~\cite{package,doi}.

\subsection{Prompt Evaluation Results}
\label{sec:development_results}

We evaluated the prompt templates for S2R identification and extraction in terms of precision, recall, and F1 score, by executing these two phases in isolation. 
The F1 score was used to rank the templates. 
The S2R mapping prompt templates were evaluated by executing \tool's S2R quality assessment phase and evaluating the resulting S2R-interaction mappings. 
Since S2R mapping is a 2-step task, we evaluated each of the prompts based on the \# and \% of \textit{hits}, defined as follows.
For the first prompt, it is the number (and proportion) of correct predictions for the presence or absence of an S2R-interaction mapping in a given screen (out of the total number of predictions). 
For the second prompt, it is the number (and proportion) of correctly identified interactions for each individual S2R (out of the total number of individual S2Rs).

\begin{table}[t!]
	\centering
	\caption{Prompt Template Performance for S2R Identification}
	\label{tab:identification_results_dev_set}
	\resizebox{\columnwidth}{!}{%
		\begin{tabular}{c|c|c|c|c|c|c}
			\hline
			\textbf{Template} & \textbf{Precision} & \textbf{Recall} & \textbf{F1} & \textbf{\#TP} & \textbf{\#FP} & \textbf{\#FN} \\ \hline
			ZS & 0.929              & 0.968           & 0.948       & 183           & 14            & 6             \\ \hline
			FS   & 0.897              & 0.963           & 0.929       & 182           & 21            & 7             \\ \hline
			CoT  & 0.915              & 0.963           & 0.938       & 182           & 17            & 7             \\ \hline
		\end{tabular}%
	}
\end{table}

\begin{table}[t!]
	\centering
	\caption{Prompt Template Performance for S2R Extraction}
	\label{tab:indiv-s2r-study-results}
	\resizebox{\columnwidth}{!}{%
		\begin{tabular}{c|c|c|c|c|c|c}
			\hline
			\textbf{Template} & \textbf{Precision} & \textbf{Recall} & \textbf{F1} & \textbf{\#TP} & \textbf{\#FP} & \textbf{\#FN} \\ \hline
			ZS              & 0.918              & 0.951           & 0.934       & 234            & 21             & 12             \\ \hline
			FS              & 0.897              & 0.951           & 0.923       & 234            & 27             & 12             \\ \hline
			CoT             & 0.810              & 0.951           & 0.875       & 234            & 55             & 12             \\ \hline
		\end{tabular}%
	}
	
\end{table}

\begin{table}[t!]
	\centering
	\caption{Prompt Performance for S2R-Interaction Mapping}
	\label{tab:indiv-s2r-matching-results}
	\resizebox{\columnwidth}{!}{%
		\begin{tabular}{c|ccc|cc}
			\hline
			\multirow{2}{*}{\textbf{Template}} & \multicolumn{3}{c|}{\textbf{1st-step template}}                                                             & \multicolumn{2}{c}{\textbf{2nd-step template}}          \\ \cline{2-6} 
			& \multicolumn{1}{c|}{\textbf{\# Predictions}} & \multicolumn{1}{c|}{\textbf{\# Hits}} & \textbf{Hit Rate} & \multicolumn{1}{c|}{\textbf{\# Hits}} & \textbf{Hit Rate} \\ \hline
			{ZS}                      & \multicolumn{1}{c|}{939}                    & \multicolumn{1}{c|}{887}              & 94.5\%            & \multicolumn{1}{c|}{30}               & 76.9\%            \\ \hline
			{FS}                      & \multicolumn{1}{c|}{970}                    & \multicolumn{1}{c|}{912}              & 94.0\%            & \multicolumn{1}{c|}{26}               & 66.7\%            \\ \hline
			{CoT}                     & \multicolumn{1}{c|}{1214}                   & \multicolumn{1}{c|}{1152}             & 94.9\%            & \multicolumn{1}{c|}{18}               & 46.2\%            \\ \hline
		\end{tabular}%
	}

\end{table}

\Cref{tab:identification_results_dev_set,tab:indiv-s2r-study-results,tab:indiv-s2r-matching-results} show the performance of the designed prompt templates for the three tasks: S2R identification, individual S2R extraction, and S2R mapping. 
Among the three templates for the S2R identification task, \textit{ZS} achieved the best performance across the three metrics having the lowest \# of FP (14) and FN~(6). Likewise, for the individual S2R extraction task, the \textit{ZS} template achieved the highest precision (0.918) with the lowest \# of FP (21), sharing the same \# of FN (12) with the other two prompts. Regarding the S2R mapping task, \tool with all three templates for the 1st-step prompt achieved a similar hit rate (94.0\% to 94.9\%) and with \textit{ZS} template for the 2nd-step prompt achieved the best hit rate of 76.9\%. 

Interestingly, although prior research has shown the superiority of \textit{CoT} prompts over \textit{ZS} and \textit{FS} prompts~\cite{hou2023large,Feng2024}, this is not the case for our tasks. Via qualitative analysis of GPT-4 responses, we observed that GPT-4 with \textit{FS} and \textit{CoT} prompts tends to include more unintended text in the responses compared to \textit{ZS} prompt which results in more false positives, \eg\ \textit{CoT} template for S2R extraction generated 55 FPs while \textit{ZS} template generated 21 FPs only. We conjecture that the long and complicated input (\eg\ bug reports can be long, and interaction information can be complicated) made the task difficult for GPT-4. Moreover, having three or four examples with reasoning made the prompts even longer.

As for all three tasks, \textit{ZS} templates outperformed the other two, we utilized the \textit{ZS} templates for implementing \tool.


\section{\tool's Evaluation Design}
\label{sec:empirical_evaluation}
\tool's evaluation has two main goals: (i) to evaluate \tool's ability to provide correct quality annotations for real bug reports, 
and (ii) to examine how well \tool can infer missing S2R information in bug reports. 
We apply \tool to a test dataset (see \Cref{sec:test_dataset}) comprising 21 bug reports, in order to provide a comparison with prior work. We aim to answer the following research questions (RQs):
\begin{itemize}
	\item \textbf{RQ$_{1}$:} How effective is \tool in generating correct S2R quality annotations?
	\item \textbf{RQ$_{2}$:} How accurately can \tool infer missing S2Rs?
\end{itemize}

\subsection{Evaluation Dataset}
\label{sec:test_dataset}

We used the bug reports (\ie\ \textit{test set}) used by Chaparro \etal~\cite{Chaparro2019}, which allow us to provide a direct comparison with their approach, \EulerC. 
This dataset contains 24 bug reports \rev{of various kinds ( crashes, UI problems, and navigation problems)} from six Android applications \rev{of different domains (web browsing, WiFi network diagnosis, finance tracking, \etc). The diverse evaluation set, separate from the development set, enabled us to assess the generalizability of the developed prompts across different bug reports.}  
We discarded three bug reports, as follows: (1) two bug reports~\cite{aard81, aard104} from the Aard Dictionary App~\cite{aardapp}, because the app version 1.4.1 is unable to load its dictionary database, and (2) one bug from Time Tracker app~\cite{atimetracker1}, because we could not generate the execution model for this app as the bug report requires a rotation action which \tool does not support.
Hence, our test set contains 21 bug reports from the original \EulerC dataset. 

Since this dataset does not contain any ground truth information for evaluating \tool, we constructed the ground truth manually.  We used the same methodology discussed in \Cref{sec:identification_data_dev,sec:extraction_data_dev} to do so for identifying S2R sentences and extracting individual S2Rs.

To construct the quality assessment ground truth, the first two authors mapped the extracted individual S2Rs to \rev{GUI} interactions manually following the methodology discussed in \Cref{sec:qa_data_dev}. App execution models for the bug reports were built by parsing execution traces collected via \CrashScope's app exploration and manual app \rev{usage}.  
One author identified the reproduction interactions on the generated data and mapped such interactions with the extracted individual S2Rs from the bug report. 
They collected the mapped interactions for each individual S2R, as well as the interactions that are required to reproduce the bug, but not reported in the bug report, \ie ground truth for missing steps. 
Each individual S2R was mapped with one or more interactions in the execution model path, as needed. Using the mapped interactions and the quality assessment model (discussed in \Cref{sec:quality_model}), they assigned quality labels to each individual S2R. 
A second author performed the same steps and validated the interactions in the reproduction scenario as well as the quality annotations.  Disagreements were resolved via discussion.

In summary, we identified 73 S2R sentences out of the 275 sentences present in the 21 bug reports with a near-perfect agreement between the two authors (98.2\% agreement rate and 0.88 Cohen's kappa \cite{Cohen}). 
From the 73 S2R sentences, we extracted 82 individual S2Rs with an agreement rate of 93.9\% between the two authors.
We discarded four individual S2Rs as they represent rotation operation and the current version of \tool does not support this operation. 
We assigned the remaining 78 individual S2Rs quality annotations (\ie\ 70 S2Rs as CS, 7 S2Rs as AS, 1 S2R as VM, and 38 S2Rs as MS). 
We identified 158 missing interactions, \ie\ missing steps for the 38 MS positions (\ie S2Rs with filled-in missing interactions). 
For constructing the annotations ground truth, the two authors agreed on 90\% of the cases. Cohen's kappa for individual S2R extraction and mapping is inapplicable since the labeling is not based on a discrete set of labels. 

\subsection{Baseline Approach}

We considered \EulerC~\cite{Chaparro2019} as the baseline approach, which also aims to assess the quality of S2Rs in a bug report. 
It identifies the S2R sentences from a bug report using deep learning techniques (\eg\ CNN~\cite{o2015introduction}, Bi-LSTM~\cite{zhou2016attention}). 
It identifies individual S2Rs via analysis of discourse patterns and assigns quality annotations by employing keyword-based mapping to app UI information.  
\rev{\EulerC and \tool generate similar quality reports, therefore we can directly compare the \tool reports to the original \EulerC reports  provided by \EulerC's replication package \cite{Chaparro2019}, 
to answer the RQs.}
\looseness=-1

\subsection{Evaluation Methodology} 

We executed \tool with the 21 bug reports on the test set,  producing the quality report for each bug report, including the quality annotations and missing steps. To answer \textbf{RQ$_{1}$}, we compared the \tool assigned quality annotations with the ground truth quality annotations. To answer \textbf{RQ$_{2}$}, we evaluated the generated missing steps by \tool against the ground truth missing steps. For both RQs, we computed precision, recall, and F1 score. We applied the same process for \EulerC and qualitatively analyzed the false positives (FP) and negatives (FN) to understand the limitations of both approaches.


\section{Results}
\label{sec:results}
\subsection{RQ$_1$: Quality Annotation Results}
	We compared all the quality annotations in the ground truth with those produced by both \tool and \EulerC.   
	\EulerC fails to identify four (among the 78) S2Rs from the ground truth, whereas \tool fails to identify two \rev{S2Rs}. We considered these in our analysis as they represent false negatives.
	
	Table \ref{tab:qa-metrics-comparison-results} compares the performance of \tool and \EulerC for each quality annotation and in aggregate across all annotations. Overall performance metrics were computed by summing the TPs, FPs, and FNs. Overall, \tool outperforms \EulerC in S2R annotation by a relative improvement of 25.2\% in terms of F1 score. \rev{The overall performance difference between \tool and \EulerC are statistically significant according to the Wilcoxon test (p-values = 0.03, 0.004, and 0.005 for precision, recall, and F1 scores, respectively).}
	\looseness=-1

	Both \rev{\tool and \EulerC} incorrectly labeled \rev{two S2Rs as CS}. For example, the S2R \textit{"Add a split"} from report \#699 from GnuCash App \cite{gnucash699} requires a user to perform \textit{tap add split button} and \textit{type split amount} interactions. However, GPT-4 \rev{identified only one mapped interaction for this S2R.}
	\EulerC also \rev{incorrectly labeled it as CS} because the matching algorithm used by \EulerC is too restrictive: it \rev{maps an S2R to} one interaction even if multiple interactions exist on the screen. \rev{Moreover, \EulerC failed to annotate 16 S2Rs as CS, while \tool failed  to annotate four S2Rs as CS.}
	
	Furthermore, \EulerC incorrectly \rev{labeled} S2Rs as AS for two S2Rs where only one \rev{mapped interaction exist}, whereas \tool never \rev{made such errors}. For example, the individual S2R, \textit{"Select an event"} from bug report \#154 in schedule-campfahrplan \cite{schedule154} is annotated as AS by \EulerC because it incorrectly \rev{mapped to multiple actions (\eg\ long click or click)}. However, \tool accurately \rev{annotated the S2R as CS}. Overall, we observed this is because (i) GPT-4 is able to correctly identify whether the S2Rs refers to single or multiple interactions, and (ii) \tool can \rev{ map an S2R to} multiple interactions on the current GUI screen when reporters \rev{combine steps or use generic verbs}. Moreover, \EulerC failed to \rev{annotate  four S2Rs as AS while \tool failed to annotate two S2Rs as AS.}
	\looseness=-1
	
	Additionally, the vocabulary mismatch (VM) annotation is typically assigned when interactions are not present in the GUIs. \EulerC produced 12 false positive VMs due to low graph coverage and its inability to infer the actual step in the app, even with the capability of adding screens/interactions to the graph while executing S2Rs (see \EulerC's paper for details~\cite{Chaparro2019}). In contrast, \tool incorrectly annotated two \rev{S2Rs} only as VM. The reasons for \EulerC failing more than \tool can be attributed to the restrictive matching algorithm \EulerC uses, whereas \tool leverages GPT-4 to map interactions to S2Rs. 
	\rev{It is also possible that \EulerC could not cover the necessary screens even after random exploration, whereas \tool utilizes a more complete graph in mapping interactions for S2Rs that traverse many GUI app screens.}
	
	\begin{table}[t]
		\centering
\vspace{-0.7em}
		\caption{Quality Annotation Results (\tool vs. \EulerC)}
	\label{tab:qa-metrics-comparison-results}
\resizebox{\columnwidth}{!}{%
	\begin{tabular}{c|c|c|c|c|c|c|c}
		\hline
		\textbf{QAs}           & \textbf{Approach} & \textbf{Precision} & \textbf{Recall} & \textbf{F1} & \textbf{\#TP} & \textbf{\#FP} & \textbf{\#FN}\\
		\hline
		\multirow{2}{*}{CS} & \EulerC & 0.964 & 0.771 & 0.857 & 54 & 2 & 16\\
		&\tool & 0.971 & 0.943 & 0.957 & 66 & 2 & 4\\
		\hline
		
		\multirow{2}{*}{AS} & \EulerC & 0.600 & 0.429 & 0.500 & 3 & 2 & 4\\
		&\tool & 1.000 & 0.714 & 0.833 & 5 & 0 & 2\\
		\hline
		
		\multirow{2}{*}{MS} & \EulerC & 0.600 & 0.553 & 0.575 & 21 & 14 & 17\\
		&\tool & 0.750 & 0.789 & 0.769 & 30 & 10 & 8\\
		\hline
		
		\multirow{2}{*}{VM} & \EulerC & 0.077 & 1.000 & 0.143 & 1 & 12 & 0\\
		&\tool & 0.333 & 1.000 & 0.500 & 1 & 2 & 0\\
		\hline
		
		\multirow{2}{*}{\textbf{Overall}} & \EulerC & \textbf{0.725} & \textbf{0.681} & \textbf{0.702} & \textbf{79} & \textbf{30} & \textbf{37}\\
		&\tool & \textbf{0.879} & \textbf{0.879} & \textbf{0.879} & \textbf{102} & \textbf{14} & \textbf{14}\\
		\hline
		
	\end{tabular}%
 }

	\end{table}
	
	\rev{There can be one or multiple missing steps  between two consecutive mapped interactions for two S2Rs. In such cases, we annotated the latter S2R as MS for analysis purposes.}
In \EulerC's 14 misclassified MS annotations, the suggested missing steps were not necessary for bug reproduction. 
In contrast, out of the ten individual S2Rs misclassified as MS by \tool, four involved completely unnecessary steps for bug reproduction, and \rev{the remaining six were due to the incorrectly mapped interactions.} 
For example, the S2R, \textit{"Select Units"} from bug report \#12 from \rev{DroidWeight \cite{droidweight12} could not map to a GUI interaction because the corresponding GUI component description is "back modal" and that GUI interaction was considered a missing step.} This leads to the necessity for developers to use meaningful GUI descriptions in the underlying code. A possible strategy to address this issue is to use GUI interaction images in the prompt to GPT-4 to provide additional context. \rev{Moreover, \tool failed to identify MS annotations for 8 S2Rs whereas \EulerC failed in 17 S2Rs, meaning \tool is more successful in identifying intermediate paths between two S2Rs.}

\subsection{RQ$_2$: Missing S2R Results}

\begin{figure}[t]
	\vspace{-1em}
	\centering
	\includegraphics[width=0.5\linewidth]{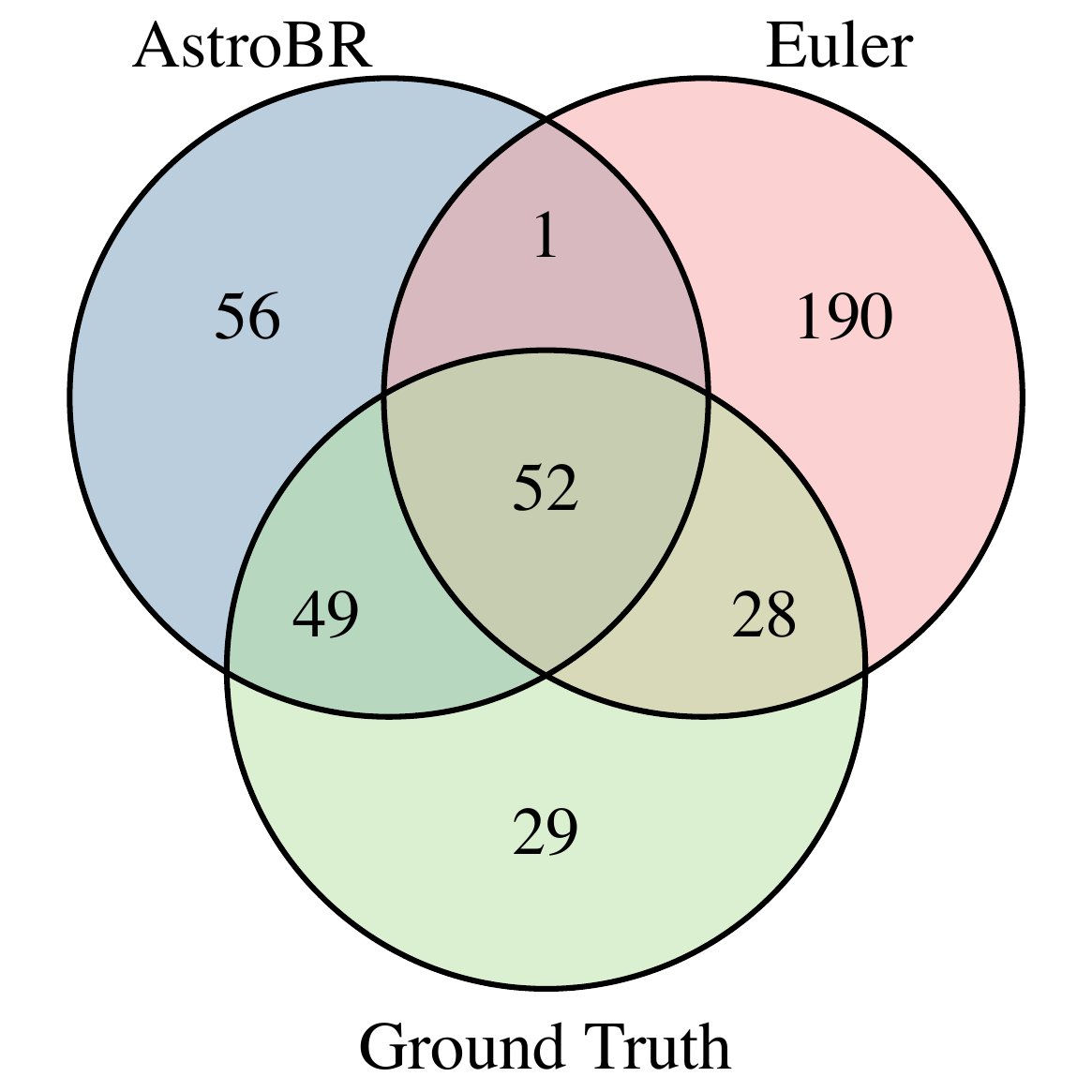}
	\caption{\# of Missing Steps Generated by \tool}
	\label{fig:ms-venn}
	\vspace{-0.3cm}
\end{figure}

In addition to evaluating \EulerC and \tool's ability to assign MS annotations, we also calculated the number of intermediate steps missing between \rev{two} consecutive S2Rs. The ground truth dataset consists of 158 missing steps (7.5 per bug report on average) across 21 bug reports. \EulerC suggested 271 missing steps (12.9 missing steps per bug report on average), and 80 of those steps are present in the ground truth, achieving a precision of 0.295, a recall of 0.506, and an F1 score of 0.373. In contrast, \tool suggested 158 missing steps (7.5 per bug report on average), 101 of which are in the ground truth, resulting in a precision, recall, and F1 score of 0.639. This means that \tool outperforms \EulerC by 71.4\% in terms of F1 score;  \tool not only produces fewer incorrect, missing steps but also identifies more \rev{accurate} missing steps.

We also analyzed if the missing steps provided by \tool are different compared to \EulerC, presented in Figure \ref{fig:ms-venn}. 

Both \EulerC and \tool successfully identified 52 correct missing steps, which represents 33\% of the missing steps in the ground truth. These steps are primarily the setup steps commonly performed in the apps; however, reporters often forgo describing such steps in bug reports. For example, nine bug reports \rev{from the Gnucash app \cite{gnucashApp} require \textit{"click the next button"} as the initial setup steps.} \rev{In addition,} both approaches produced one unnecessary missing step because of app traversal in an incorrect path. Moreover, both \tool and \EulerC failed to identify 29 crucial steps that reporters often ignore. These types of steps include: (i) actions in a popup dialog (\eg\ \textit{"Click OK button"}), and (ii) implicit steps that are easy to understand for humans but challenging for an automated tool to infer. \rev{For example, the S2R, \textit{"Click on existing transaction"} from Gnucash's bug report \#699~\cite{gnucash699}  requires the existence of a transaction in the app before clicking on the  GUI component corresponding to the transaction, but that S2R is not explicitly described in the bug report.}

\EulerC identified 28 ground truth missing steps that \tool failed to detect. This occurred for two main reasons. First, \tool \rev{ would sometimes perform incorrect mapping between S2Rs and GUI interactions due to choosing an alternate path to reach from the current screen to a mapped screen, thus skipping crucial steps.} For example, the S2R \textit{"Navigate to Service Intervals screen"} from android-mileage's bug report \#65~\cite{mileage65} was incorrectly mapped \rev{to} \textit{"click the ok button"}, which lead to omitting the necessary step \textit{"open the app menu"}. Second, \EulerC identifies more initial setup steps required on a screen to proceed from the current GUI screen to the next due to its random exploration; however, \tool often ignores such steps, as it identifies the minimal steps necessary to perform on the current screen. To resolve this problem, future versions of \tool may incorporate knowledge of specific app functionalities \rev{to suggest missing} steps in bug reports. 

\rev{Across 21 bug reports, \EulerC produces 190 unnecessary steps that are not detected by \tool, whereas \tool only produces 56 such steps, excluding the ones detected by \EulerC.} 
Although \EulerC prioritizes recall over precision to ensure the presence of necessary missing steps from which reporters can choose, the large number of missing steps may confuse developers when reproducing the bug.
The reason for \rev{the large number of unnecessary steps} is attributed to the incorrect S2R interaction matches in the GUIs. However, \EulerC's random exploration strategy in the graph exacerbates this problem by producing an excessive number of unnecessary steps.

\tool correctly identified 49 missing steps in the ground truth test dataset that \EulerC failed to detect. There are primarily two reasons: (i) \EulerC includes \textit{"Open App"} step only if explicitly described in the bug report, yet only two out of 21 bug reports contain this step---\tool includes this step in all \rev{quality} reports; and (ii) \EulerC's graph prevents the identification of GUI information for navigation components such as the navigation drawer and input widgets such as the spinner. \rev{For example, in Gnucash's bug report \#615~\cite{gnucash615}, \EulerC identifies the step \textit{"Tap the `Navigation drawer opened' image button"} but fails to identify the interaction for the \textit{"Manage Books"} that becomes visible when the navigation drawer is opened.} The graph used in \tool correctly identifies such interaction, exhibiting its ability to handle complicated GUI structures. 
\looseness=-1


\section{Related Work}
\label{sec:related_work}

\rev{Researchers have investigated bug reports for a variety of purposes including bug report management~\cite{saha2024toward,Adnan:msr25,zou2018practitioners,Mahmud:ICSE2024}, understanding bug resolution~\cite{Saha:icse25}, predicting bug priority and severity~\cite{umer2019cnn,tian2015automated,huang2022bug}, categorizing bug types~\cite{somasundaram2012automatic,catolino2019not},  identifying duplicate bugs~\cite{yan2024semantic,Zhou2012a,he2020duplicate,cooper2021takes,chaparro2019reformulating,chaparro2016vocabulary}, reproducing bugs~\cite{Fazzini2018,Zhao2019,Feng2024,Wang2024,bernal2020translating,bernal2023translating,Havranek2021}, and localizing buggy code~\cite{florez2021combining,chaparro2019using,chaparro2017using,chaparro2016reduction}. We discuss the most  closely related work in this section.}

\textbf{Assessing Bug Report Quality.} 
Past research in assessing the quality of bug reports is primarily focused on the readability, coherence, and inclusion of the necessary components within bug reports. Zimmermann \cite{Zimmermann2010} proposed an approach to assess the quality of bug reports by classifying them as bad, neutral, or good, considering various features such as keyword completeness, patches, screenshots, and readability.
Dit  \etal~\cite{Dit2008} evaluates the quality of bug reports based on the coherence of comments in bug report discussions.  
Linstead  \etal \cite{Linstead2009} later proposed a different textual coherence calculation technique, utilizing an information-theory-based approach by measuring the entropy of the distribution of latent topics in bug reports. 
Very recently, Bo  \etal introduced ChatBR~\cite{Bo2024}, which assesses and generates S2Rs if absent but does not evaluate generated S2R quality. \tool advances upon ChatBR by assessing S2Rs using annotations by determining whether S2Rs can be mapped to application UI interactions.
\looseness=-1

Chaparro  \etal \cite{Chaparro2019} introduced \EulerC, an approach that provides quality annotation for the S2Rs in bug reports.
\EulerC is the closest related work, in as much as they produce the same type of quality reports, given a bug description.
Unlike \EulerC, \tool uses LLMs to generate the quality reports. In consequence, its internals are fundamentally different, particularly in how the app model is explored. These improvements lead to more effective quality annotations as illustrated by the results of \tool's evaluation.  

\textbf{Automated Bug Reproduction.} 
Researchers introduced various techniques to generate test cases for automated bug reproduction to diagnose, validate, and understand bugs. Fazzini  \etal \cite{Fazzini2018} developed Yakusu, which combines program analysis and text processing techniques to create test cases for bug reproduction.  
Zhao  \etal \cite{Zhao2019} proposed ReCDroid, to reproduce crashes. ReCDroid formulates a dynamic ordered event tree (DOET) leveraging GUI components and event transitions, which aids in traversing GUIs for a given app and prioritizes relevant GUI components for exploration. 
Feng and Chen~\cite{Feng2024} introduced AdbGPT, which focuses on automatically reproducing bugs using LLMs. AdbGPT extracts actions and objects from S2Rs through prompt engineering and later leverages GUI encoding and LLMs to replay bugs within app screens. Wang  \etal \cite{Wang2024} proposed ReBL that mitigates different limitations of AdbGPT and utilizes the entire bug report instead of only using S2Rs to improve the contextual reasoning of the LLMs in automatically reproducing bugs.

\textbf{Interactive Bug Reporting Systems.}
Researchers have proposed systems for interactive bug reporting, which typically aim also at improving the quality of the bug reports.
 Moran  \etal proposed \Fusion \cite{Moran2015} that allows reporters to choose available actions and GUI components from dropdown lists, resulting in more structured and comprehensive bug reports. 
 Fazzini  \etal proposed \ebug \cite{Fazzini:TSE22} that extends \Fusion and suggests potential S2Rs to the reporters alongside the dropdown lists available to \Fusion.
 Song  \etal proposed \burt \cite{song2022toward, song2022burt}, a chatbot that guides the reported and verifies the quality of bug information in real-time, providing suggestions to the reporters.


\section{Threats to Validity and Limitations}
\label{sec:threats}
\textbf{Construct Validity.}
The main threats to construct validity stem from manually verifying the matching of the interactions extracted from the S2R sentences to the information on the execution model and constructing a ground truth dataset. 
To mitigate this threat, two authors independently carried out the manual verification tasks and ground truth creation, following well-defined and replicable methodologies.
More so, we computed and reported agreement levels, which are very high in all cases. 
\looseness=-1

\textbf{Internal Validity.}
Selecting the optimal prompt can be challenging for any use of GPT-4, let alone for multiple distinct tasks, and this process of finding the best prompt impacts the performance of our approach.
We selected the best prompt by evaluating 14 prompt templates, using three prompting strategies (\ie zero-shot, few-shot, and chain-of-thought) on a rich development set of bug reports from multiple applications.
\looseness=-1 

\textbf{External Validity.}
Our results are compared with the state-of-the-art, \EulerC, where we used 21 bug reports from their original dataset across six applications. We could not increase the dataset size for comparison due to difficulties in running the \EulerC tool. However, our approach is built by analyzing a dataset with bug reports from nine different applications consisting of four types of bugs. Therefore, \tool can be generalized to diverse types of bug reports.
\rev{Moreover, AstroBR currently supports the most frequently used GUI interactions in Android applications (tap, long tap, \etc). While the lack of support for certain types of interactions (\eg rotation) is a limitation, this is not due to the inherent design of the approach, and the support of these features can be added through additional engineering effort in future work.}

\textbf{Limitations.}
\tool's performance depends on the completeness of the app execution model.
The automated execution information collected with \CrashScope may result in an incomplete execution model.
To overcome this issue, we collected information from manual app executions. 


\section{Conclusions and Future Work}
\label{sec:conclusions}

Providing quality feedback to bug reporters at reporting time promises to result in bug reports that are easier to understand and reproduce by developers.  
We found that using LLMs (\ie GPT-4) for automatically extracting and analyzing S2Rs from natural language bug reports, and matching them to GUI interactions, is very effective, resulting in better quality annotations than state-of-the-art approaches.
As with any other applications of LLMs, their performance is highly dependent on the prompting quality.
By investigating 12 prompt templates, using different prompting strategies, we observed that the use of GPT-4 in this context is quite robust with respect to the type of prompt used.
That is, different prompt templates lead to similar performance levels.

Future work will focus on expanding the evaluation to larger data sets and tackle the quality of other elements of bug reports, such as, the observed and expected behavior. \rev{We will also add support for additional types of interactions (\eg\ rotation) and perform a user study to assess \tool's usability by engineers in real-world scenarios. In addition, we will compare multiple LLMs in a follow-up extension of this work.}


\section*{Acknowledgments}
This work is supported by U.S. NSF grants CCF-1955853, CCF-2343057, and CCF-2441355. The opinions, findings, and conclusions expressed in this paper are those of the authors and do not necessarily reflect the sponsors' opinions.

\balance

\bibliographystyle{IEEEtran}
\bibliography{references}

\end{document}